\documentclass[conference]{IEEEtran}
\IEEEoverridecommandlockouts
\usepackage[
    backend=biber,
    maxcitenames=3,
    minbibnames=1,
    mincitenames=1,
    maxbibnames=3,
    uniquelist=false,
    style=ieee,
    url=false,
    doi=true,
    eprint=false,
    isbn=false,
    ]{biblatex}
\AtBeginBibliography{\footnotesize}
\usepackage{amsmath,amssymb,amsfonts}
\usepackage{algorithmic}
\usepackage{graphicx}
\usepackage{textcomp}
\usepackage{xcolor}

\usepackage{tabularray}
\UseTblrLibrary{booktabs}
\usepackage[nohyperlinks, printonlyused, withpage, smaller, nolist]{acronym}
\usepackage{listings} % listings
\usepackage[inline]{enumitem}
\usepackage[colorlinks=true,urlcolor=teal,
            linkcolor=teal,citecolor=teal]{hyperref} % urls and hyperlinks
\usepackage{tikz}   
\usetikzlibrary{quantikz2,positioning}
\usepackage[T1]{fontenc}    % use 8-bit T1 fonts
\usepackage{csquotes}
\usepackage{xspace}
\usepackage{censor}
\usepackage{orcidlink}

\newboolean{anonymous}
\setboolean{anonymous}{false}
\begin{acronym}[TROLL]
    \acro{nisq}[NISQ]{Noisy Intermediate Scale Quantum}
    \acro{ftqc}[FTQC]{Fault-Tolerant Quantum Computing}
    \acro{vqc}[VQC]{Variational Quantum Circuit}
    \acro{qml}[QML]{Quantum Machine Learning}
    \acro{qfm}[QFM]{Quantum Fourier Model}
    \acro{qc}[QC]{Quantum Computing}
    \acro{pqc}[PQC]{Parameterized Quantum Circuit}
    \acro{ml}[ML]{Machine Learning}
    \acro{spam}[SPAM]{State Preparation and Measurement}
    \acro{pd}[PD]{Phase Damping}
    \acro{bf}[BF]{Bit Flip}
    \acro{pf}[PF]{Phase Flip}
    \acro{ad}[AD]{Amplitude Damping}
    \acro{dp}[DP]{Depolarization}
    \acro{kl}[KL]{Kullback-Leibler}
    \acro{fft}[FFT]{Fast Fourier Transform}
\end{acronym}

\newcommand{\etal}{\emph{et al.}\xspace}

\makeatletter
\newcommand{\linebreakand}{%
  \end{@IEEEauthorhalign}
  \hfill\mbox{}\par
  \mbox{}\hfill\begin{@IEEEauthorhalign}
}
\makeatother

\newlist{todolist}{itemize}{2}
\setlist[todolist]{label=$\square$}
\usepackage{pifont}

\ifbool{anonymous}{}{}
\ifbool{anonymous}{}{\renewcommand{\blackout}[1]{#1}}
\ifbool{anonymous}{\newcommand{\genemail}[1]{\href{mailto:xxx.xxx@xx.xx}{\blackout{#1}}}}{\newcommand{\genemail}[1]{\href{mailto:#1}{#1}}}
\ifbool{anonymous}{\newcommand{\geneurl}[1]{\href{https://thiscatexists.com/}{\blackout{#1}}}}{\newcommand{\geneurl}[1]{\url{#1}}}
\ifbool{anonymous}{\newcommand{\qmlessentials}{\blackout{QML Essentials}\xspace}}{\newcommand{\qmlessentials}{\textsc{QML-Essentials}\xspace}}

\DeclareFixedFont{\ttb}{T1}{txtt}{bx}{n}{9} % for bold
\DeclareFixedFont{\ttm}{T1}{txtt}{m}{n}{9}  % for normal
\DeclareFixedFont{\ttmtable}{T1}{txtt}{m}{n}{8}  % for normal
\newcommand\pythonstyle{\lstset{
    language=Python,
    basicstyle=\ttm,
    morekeywords={self},              % Add keywords here
    keywordstyle=\ttb\color{deepblue},
    emph={MyClass,__init__},          % Custom highlighting
    emphstyle=\ttb\color{deepred},    % Custom highlighting style
    stringstyle=\color{deepgreen},
    frame=tb,                         % Any extra options here
    showstringspaces=false
  }}
  
\newcommand\pythonstyletable{\lstset{
    language=Python,
    basicstyle=\ttmtable,
    morekeywords={self},              % Add keywords here
    keywordstyle=\ttb\color{deepblue},
    emph={MyClass,__init__},          % Custom highlighting
    emphstyle=\ttb\color{deepred},    % Custom highlighting style
    stringstyle=\color{deepgreen},
    frame=tb,                         % Any extra options here
    showstringspaces=false
  }}
\newcommand\pythoninline[1]{{\pythonstyle\lstinline!#1!}}
\newcommand\pythoninlinetable[1]{{\pythonstyletable\lstinline!#1!}}

\setlist[enumerate]{label=({\arabic*})}

\begin{document}

\title{\qmlessentials---A Framework for working with Quantum Fourier Models\\}

% Paper for QSW conference; details here: https://services.conferences.computer.org/2025/info-for-authors/

\author{
    \IEEEauthorblockN{\blackout{Melvin Strobl}\textsuperscript{*} \orcidlink{0000-0003-0229-9897}}
    \IEEEauthorblockA{\blackout{\textit{Karlsruhe Institute of Technology}}\\
        \blackout{Karlsruhe, Germany} \\
        \genemail{melvin.strobl@kit.edu}}
    \and
    \IEEEauthorblockN{\blackout{Maja Franz}\textsuperscript{*} \orcidlink{0000-0002-2801-7192}}
    \IEEEauthorblockA{\blackout{\textit{Technical University of}}\\
        \blackout{\textit{Applied Sciences Regensburg}} \\
        \blackout{Regensburg, Germany} \\
        \genemail{maja.franz@othr.de}}
    \and
    \IEEEauthorblockN{\blackout{Eileen Kuehn} \orcidlink{0000-0002-8034-8837}}
    \IEEEauthorblockA{\blackout{\textit{Karlsruhe Institute of Technology}}\\
        \blackout{Karlsruhe, Germany} \\
        \genemail{eileen.kuehn@kit.edu}}
    \linebreakand
    \IEEEauthorblockN{\blackout{Wolfgang Mauerer} \orcidlink{0000-0002-9765-8313}}
    \IEEEauthorblockA{\blackout{\textit{Technical University of}}\\
        \blackout{\textit{Applied Sciences Regensburg}}\\
        \blackout{\textit{Siemens AG, Technology}}\\
        \blackout{Regensburg/Munich, Germany}\\
        \genemail{wolfgang.mauerer@othr.de}}
    \and
    \IEEEauthorblockN{\blackout{Achim Streit} \orcidlink{0000-0002-5065-469X}}
    \IEEEauthorblockA{\blackout{\textit{Karlsruhe Institute of Technology}}\\
        \blackout{Karlsruhe, Germany} \\
        \genemail{achim.streit@kit.edu}}
}

\maketitle
\begingroup\renewcommand\thefootnote{*}
\footnotetext{Equal contribution}
\endgroup

\begin{abstract}
    In this work, we propose a framework in the form of a Python package, specifically designed for the analysis of \acl{qml} models.
    This framework is based on the PennyLane simulator and facilitates the evaluation and training of \aclp{vqc}.
    It provides additional functionality ranging from the ability to add different types of noise to the classical simulation, over different parameter initialisation strategies, to the calculation of expressibility and entanglement for a given model.
    As an intrinsic property of \aclp{qfm}, it provides two methods for calculating the corresponding Fourier spectrum: one via the Fast Fourier Transform and another analytical method based on the expansion of the expectation value using trigonometric polynomials.
    It also provides a set of predefined approaches that allow a fast and straightforward implementation of \acl{qml} models.
    With this framework, we extend the PennyLane simulator with a set of tools that allow researchers a more convenient start with \aclp{qfm} and aim to unify the analysis of \aclp{vqc}.
\end{abstract}


\acresetall

\section{Introduction}

\ac{vqc} are a promising approach on the \ac{nisq} era towards \ac{ftqc}.
With pervasive challenges such as barren plateaus~\cite{ragone_lie_2024} and the resulting limitations on trainability, there are still many questions to be answered~\cite{zimboras_myths_2025}.
It was found that \acp{vqc} following a certain structure can be represented by a truncated Fourier series~\cite{schuld_effect_2021}.
In the remainder of this paper we will refer to such models as \ac{qfm} (which are different from the well-known Quantum Fourier Transform).
The Fourier spectrum is considered as an important property to characterise a \ac{qml} model, as it reflects its ability to learn non-linearities with asymptotically being a universal function approximator~\cite{schuld_effect_2021}.
The study of the spectrum of \acp{qfm} sparked a series of exciting research~\cite{wiedmann_fourier_2024,jaderberg_let_2024,nemkov_fourier_2023,kyriienko_generalized_2021} and gives insights on the dequantisability of a given \ac{qml} model~\cite{sweke_potential_2025}, also in the context of noisy computation~\cite{fontana_spectral_2022}.

By introducing \qmlessentials we provide a tool to explore the properties of \acp{qfm} in the context of \ac{qml}.
We also implement algorithms to compute key metrics of \acp{vqc}, such as expressibility and entangling capability~\cite{sim_expressibility_2019}.
Since noise is an intrinsic property of \ac{qc}, \qmlessentials provides the ability to apply noise to the simulation to investigate the behaviour of such algorithms on today's \ac{nisq} devices.
By providing open access to the algorithms, we aim to standardise the computation of these metrics for specific algorithms across the \ac{qml} community~\cite{Mauerer:2022}.

In this article, we first discuss related work and other quantum frameworks for working with \ac{qml} models in~\autoref{sec:rel_work}.
We then give an overview of the framework, the implemented algorithms and metrics in~\autoref{sec:framework}.
In~\autoref{sec:examples}, we show some examples of how the framework can be used to reproduce existing results in the literature and conclude in~\autoref{sec:conclusion}.

\section{Related Work}
\label{sec:rel_work}

We acknowledge that there are numerous other exceptional software projects that facilitate research in the field of \ac{qc}~\cite{Carbonelli:2024}.

The PennyLane~\cite{bergholm_pennylane_2022} library, which functions as a backend for the proposed framework, offers a classical simulator of quantum computing and a comprehensive suite of tools that extend the capabilities of the simulator.
These include the ability to simulate noise, calculate gradients, and perform numerous other operations.
Nevertheless, while PennyLane also provides an implementation to calculate the Fourier coefficients of a given circuit using the \ac{fft}, it is too limited for extensive research with \acp{qfm}.

With PennyLane being more focused on hybrid computation and machine learning, the framework Qiskit~\cite{javadi-abhari_quantum_2024} is targeted towards more hardware-aware development.

In comparison to PennyLane, Qiskit offers a more extensive suite of tools for direct manipulation of quantum circuits during the trans\-pi\-la\-tion process, as well as for direct manipulation at the pulse level.
However, it should be noted that Qiskit provides limited support for \ac{qml} models compared to PennyLane, which restricts its applicability to \acp{qfm}.

Horqrux, a backend of Qadence~\cite{seitz_qadence_2025}, is a simulation framework designed for \ac{qml}. It provides options to fit non-linear functions as well as solving partial differential equations.
There are other full-stack frameworks such as Qibo~\cite{efthymiou_qibo_2022} and QRISP~\cite{seidel_qrisp_2024} that provide different simulator backends and allow for working with \ac{qml} in general to some extent, but are not tailored to \ac{qfm}.

To the best of our knowledge, no framework unifies the analysis of \acp{qfm} in the context of \ac{qml} with the variety of tools that are elaborated in this paper.

\section{\qmlessentials Framework}
\label{sec:framework}

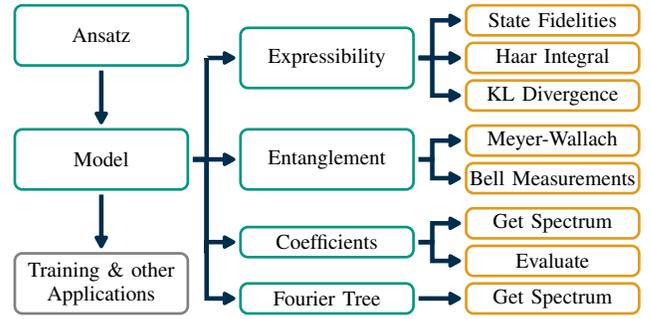
\begin{figure}[htb]
    \centering
    \begin{tikzpicture}
        % Colors
        \definecolor{kitgreen}{RGB}{0, 150, 130}  %009682
        \definecolor{kitorange}{RGB}{223, 155, 27} %DF9B1B
        \definecolor{kitdarkblue}{RGB}{0, 45, 75} %002D4C
        \tikzstyle{box1}=[draw, rounded corners=1mm, line width=1.0pt,
        text width=23mm, minimum height=4mm,
        inner sep=0pt,
        draw=kitgreen,
        font=\footnotesize, align=flush center
        ]
        \tikzstyle{box11}=[draw, rounded corners=1mm, line width=1.0pt,
        text width=23mm, minimum height=8mm,
        inner sep=0pt,
        draw=kitgreen,
        font=\footnotesize, align=flush center
        ]
        \tikzstyle{box2}=[draw, rounded corners=1mm, line width=1.0pt,
        text width=23mm, minimum height=8mm,
        inner sep=0pt,
        draw=gray,
        font=\footnotesize, align=flush center
        ]
        \tikzstyle{box3}=[draw, rounded corners=1mm, line width=1.0pt,
        text width=23mm, minimum height=4mm,
        inner sep=0pt,
        draw=kitorange,
        font=\footnotesize, align=flush center,
        ]
        \tikzstyle{arr}=[-{Triangle[length=2mm,width=2mm,round]}, line width=0.6mm, kitdarkblue, shorten >=0.5mm, shorten <=0.5mm]

        % Nodes
        \node[box11] (ansatz) at (0,3.5) {Ansatz};
        \node[box11] (model) at (0,1.85) {Model};
        \node[box2] (training) at (0,0.2) {Training \& other Applications};

        \node[box11] (expressibility) at (3,3.2) {Expressibility};
        \node[box11] (entanglement) at (3,1.85) {Entanglement};
        \node[box1] (coefficients) at (3,0.75) {Coefficients};
        \node[box1] (fourier) at (3,0) {Fourier Tree};

        \node[box3] (state) at (6,3.7) {State Fidelities};
        \node[box3] (haar) at (6,3.2) {Haar Integral};
        \node[box3] (kl) at (6,2.7) {KL Divergence};
        \node[box3] (meyer) at (6,2.1) {Meyer-Wallach};
        \node[box3] (bell) at (6,1.6) {Bell Measurements};
        \node[box3] (spectrum1) at (6,1) {Get Spectrum};
        \node[box3] (spectrum2) at (6,0.5) {Evaluate};
        \node[box3] (spectrum3) at (6,0) {Get Spectrum};

        % Edges
        \draw[arr] (ansatz) -- (model);
        \draw[arr] (model) -- (training);
        \draw[arr] (model) -| +(1.4,1.35) |- (expressibility);
        \draw[arr] (model) --  (entanglement);
        \draw[arr] (model) -| +(1.4,-1.25) |-  (coefficients);
        \draw[arr] (model) -| +(1.4,-1.85) |-  (fourier);

        \draw[arr] (expressibility) -| +(1.4,0.5) |- (state);
        \draw[arr] (expressibility) -- (haar);
        \draw[arr] (expressibility) -| +(1.4,-0.5) |- (kl);

        \draw[arr] (entanglement) -| +(1.4,0.25) |- (meyer);
        \draw[arr] (entanglement) -| +(1.4,-0.25) |- (bell);

        \draw[arr] (coefficients) -| +(1.4,0.25) |- (spectrum1);
        \draw[arr] (coefficients) -| +(1.4,-0.25) |- (spectrum2);
        \draw[arr] (fourier) -- (spectrum3);
    \end{tikzpicture}

    \caption{Overview of the classes of the \qmlessentials framework. Modules of the framework are depicted in teal, an excerpt of their corresponding functions in orange.}
    \label{fig:qml_essentials_ov}
\end{figure}

The modules of the Python package, which is available on PyPi\footnote{\geneurl{https://pypi.org/project/qml-essentials/}} and Github\footnote{\geneurl{https://github.com/cirKITers/qml-essentials/}}, are summarised in~\autoref{tab:modules} with an overview of its dependencies  shown in~\autoref{fig:qml_essentials_ov}.
The \pythoninline{Model} sits at the core of \qmlessentials and is built upon a chosen Ansatz.
It can either be used in training and other applications outside of \qmlessentials, or passed to the corresponding functions of the \pythoninline{Expressibility} (\autoref{par:expressibility}), \pythoninline{Entanglement} (\autoref{par:entangling_capability_mw} and~\autoref{par:entangling_capability_bell}), \pythoninline{Coefficients} (\autoref{par:coefficients_fft}) and \pythoninline{FourierTree} (\autoref{par:coefficients_analytical}) modules explained in the referenced sections.
In this work, we only focus on the aforementioned aspects of our framework.
Nonetheless it should be noted that there is a plethora of other features in \qmlessentials such as
\begin{enumerate*}[label=(\arabic*)]
    \item different initialisation strategies and parameter sampling,
    \item changing of, or providing custom feature maps,
    \item different measurements based on provided output shapes,
    \item caching of results using hashed parameters and
    \item oversampling of the Fourier spectrum.
\end{enumerate*}
These features are not covered in this article but are explained on our curated documentation page.\footnote{\geneurl{https://cirkiters.github.io/qml-essentials/}}

\begin{table}[htb]
    \centering
    \caption{Python module overview.}
    \label{tab:modules}
    \begin{tblr}{width=\linewidth,
            colspec={lX[l]},
            row{2-Z}={belowsep=0.15em},
            booktabs}
        \toprule
        \textbf{Module}                    & \textbf{Description}                                                                                                        \\
        \midrule
        \pythoninlinetable{Model}          & Data-reuploading model class with various options to change initialisation strategy, encoding, measurement qubit and noise. \\
        \pythoninlinetable{Ansaetze}       & Set of circuits to be used within a model.                                                                                  \\
        \pythoninlinetable{Coefficients}   &                                                                                                                             %Calculation of coefficients using either \ac{fft} or analytical method.\\
        Calculate coefficients (\ac{fft} or analytical).                                                                                                                 \\
        \pythoninlinetable{Expressibility} & Tools to calculate expressibility.                                                                                          \\
        \pythoninlinetable{Entanglement}   & Calculate entangling capability (Meyer-Wallach measure or Bell measurements).                                               \\
        \bottomrule
    \end{tblr}
\end{table}

In the following, we introduce the main concepts to \ac{qml} models, and explain how they can be  utilised and analysed using modules of the \qmlessentials framework.

\subsection{Quantum Machine Learning}

Generally, the goal of \ac{qml} is the same as for classical machine learning, that is to \enquote{learn} a function $f$, using a function approximator $f_\theta$, parametrised by $p$ parameters $\theta \in \mathbb{R}^p$.
The parameters $\theta$ are \enquote{trained} to minimise the difference between $f(x)$ and $f_\theta(x)$, for some input $x \in \mathbb{R}^N$ with $N$ input features, using classical optimisation routines, such as gradient descent~\cite{schuld_supervised_2018}.

In the case of \ac{qml}, the function approximator utilises a \ac{vqc}, characterised by the parametrised unitary $U_\theta$, acting on a system of $n$ qubits.
The function evaluation of the approximator then corresponds to the expectation value of an observable $\mathcal{O}$ on the circuit's state:
\begin{equation}
    f_{\theta}(x)=\bra{0}^{\otimes n} U_\theta^{\dagger}(x) \mathcal{O} U_\theta(x) \ket{0}^{\otimes n}.
    \label{eq:expval_pqc}
\end{equation}

\subsection{Quantum Fourier Models}

The basis of our framework forms the \ac{qfm}, which utilises a specific structure of unitary $U_\theta$, namely an interleaving pattern of trainable- and encoding unitaries $W \coloneq W_\theta$ and $S$ respectively:
\begin{equation}
    U_\theta(x)=W^{(L+1)} S(x) W^{(L)} \cdots W^{(2)} S(x) W^{(1)}.
\end{equation}

As shown in the seminal works of Ref.~\cite{perez-salinas_data_2020} and Ref~\cite{schuld_effect_2021}, circuits following this unitary structure
\begin{enumerate*}[label=(\arabic*)]
    \item represent universal function approximators~\cite{perez-salinas_data_2020} and
    \item can be represented as a truncated Fourier series with the frequencies $\omega \in \Omega$ and their corresponding magnitudes $c_{\omega}(\theta)$ as its characteristic properties:
\end{enumerate*}

\begin{equation}
    f_{\theta}(x)=\sum_{\omega \in \Omega} c_{\omega}(\theta) e^{i \omega x}.
    \label{eq:expval_qfm}
\end{equation}

\subsubsection{Coefficient calculation using the \acs{fft}}
\label{par:coefficients_fft}

Practically, basic signal analysis allows us to retrieve the coefficients of a given circuit by evaluating its expectation value~\autoref{eq:expval_pqc} at different inputs and applying a \ac{fft}.
Given a model, its coefficients can be estimated using the static \pythoninline{Coefficients.get_spectrum()} method.

While providing a fast and in general reliable method, the estimation of coefficients using the \ac{fft} also bears some disadvantages.
Firstly, the frequencies must be evenly spaced or the number of sampling points must be chosen such that intermediate frequencies are captured correctly.
Even then it only gives an approximation based on the provided sampling points without taking into account the actual circuit properties.
This can lead to scenarios in which frequencies are not captured at all~\cite{wiedmann_fourier_2024}.
Secondly, the highest frequency must be estimated in advance to fulfil the Nyquist criterium.
Otherwise frequency artefacts caused by the repeating structure of the Fourier transform will be observed.

\subsubsection{Coefficient calculation using the analytical method}
\label{par:coefficients_analytical}

Wiedmann~\etal~\cite{wiedmann_fourier_2024} introduced an analytical method to estimate the coefficients of a given \ac{qml} model.
The proposed algorithm builds upon the work by Nemkov~\etal~\cite{nemkov_fourier_2023}, which proposes an expansion of $f_{\theta}(x)$ in terms of trigonometric polynomials.
This method relies on all operations in the circuit being either Pauli-rotation or Clifford gates, to which any \ac{vqc} can be decomposed into.
The circuit $U_\theta$ is then transformed into a circuit that only consists of Pauli rotation gates, with all Clifford gates moved towards the end of the circuit to be included in the observable.
We implement their method in our framework and it is accessible via the \pythoninline{FourierTree.get_spectrum()} method after instantiating the \pythoninline{FourierTree} class using a \ac{qml} model.
While significantly slower, it provides an accurate estimation of the Fourier spectrum which leaves the choice between these two methods up to the user.

\subsection{Expressibility and Entangling Capability}

As the design of Ansätze in \ac{qml} is still an open field of research, our framework provides tools to calculate the expressibility and entangling capability as two common metrics of interest.

\subsubsection{Expressibility}
\label{par:expressibility}

For the expressibility we utilise the definition introduced in Ref.~\cite{sim_expressibility_2019} which is the \ac{kl} divergence~\cite{kullback_information_1951} between the distributions obtained by sampling from the Haar integral $\int_{\text {Haar }}(|\psi\rangle\langle\psi|)^{\otimes t} d \psi$ and the model $\int_{\boldsymbol{\Theta}}\left(\left|\psi_{\boldsymbol{\theta}}\right\rangle\left\langle\psi_{\boldsymbol{\theta}}\right|\right)^{\otimes t} d \boldsymbol{\theta}$ respectively:
\begin{equation}
    D_{\mathrm{KL}}\left(\hat{P}_{\text{Model}}(F ; \boldsymbol{\theta}) \| P_{\text {Haar }}(F)\right)
\end{equation}

Here, the fidelity $F=\left|\left\langle\psi_{\boldsymbol{\varphi}} \mid \psi_{\boldsymbol{\phi}}\right\rangle\right|$ is the probability of state overlaps, whereas the distributions of state overlaps is $p\left(F=\left|\left\langle\psi_{\boldsymbol{\varphi}} \mid \psi_{\boldsymbol{\phi}}\right\rangle\right|\right)$.

This metric yields zero if $\hat{P}_{\text{Model}}(F ; \boldsymbol{\theta}) = P_{\text {Haar }}(F)$, meaning the states sampled from the \ac{qfm} are Haar distributed.
For the least expressive case, i.e. the idle circuit, the KL divergence becomes $\ln(n_\text{bins})$ where $n_\text{bins}$ describes the number of bins that are used for discretising the probability distribution using a histogram.
In this work, we refer to the expressibility as the inverse of \ac{kl} divergence.

In \qmlessentials, the expressibility of a given model can be calculated with the following steps:
\begin{enumerate*}
    \item The state fidelities of two uniformly random parameter sets are calculated using the corresponding method providing a number of samples,
    \item the Haar integral for the same number of qubits as in the model is computed,
    \item the results of both calculations are passed to the \pythoninline{kullback_leibler_divergence} method of the \pythoninline{Expressibility} class to obtain the distance between the two distributions given a number of bins.
\end{enumerate*}

\subsubsection{Entangling Capability using Meyer-Wallach measure}
\label{par:entangling_capability_mw}

As shown in Refs.~\cite{meyer_global_2002, brennen_observable_2003}, the entangling capability for a $n$-qubit system can be defined based on the trace of the squared partial density matrix $\rho_k$ for subsystem $k$:

\begin{equation}
    Q(|\psi\rangle)=2\left(1-1 / n \sum_{k=0}^{n-1} \operatorname{Tr}\left[\rho_{k}^{2}\right]\right)
    \label{eq:entangling_capability_brennen}
\end{equation}

This metric has the property that if $\operatorname{Tr}\left[\rho_{j}^{2}\right]=1 \quad \forall j$, implying $Q=0$, then $\ket{\psi}$ is a product state whereas $Q=1 \iff \operatorname{Tr}\left[\rho_{k}^{2}\right]=1 / 2 \quad \forall k$, meaning the state is maximally mixed.
Notable, access to the density matrix is required for this metric, which is not available on real devices without further ado.

In \qmlessentials this metric can be calculated using the \pythoninline{meyer_wallach} method within the \pythoninline{Entanglement} module class, given a number of parameter samples.

\subsubsection{Entangling Capability using Bell-Measurements}
\label{par:entangling_capability_bell}

\begin{figure}
    \centering
    \begin{tikzpicture}
        \node[scale=0.85] {
            \begin{quantikz}[row sep=2mm, column sep=2mm]
                \lstick{\ket{0}} & \gate[2]{U_\theta} & \ctrl{2} & & \gate{H} & \meter{} \\
                \lstick{\ket{0}} & & & \ctrl{2} & \gate{H} & \meter{} \\
                \lstick{\ket{0}} & \gate[2]{U_\theta} & \targ{} & & & \meter{} \\
                \lstick{\ket{0}} & & & \targ{} & & \meter{}
            \end{quantikz}
        };
    \end{tikzpicture}
    \caption{Setup of a \enquote{Bell-Measurement} for a $2$ qubit circuit described by $U_\theta$.}
    \label{fig:bell_measurement}
\end{figure}
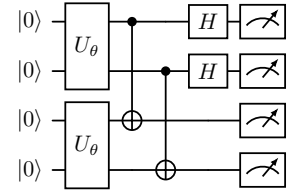

As an alternative method to calculate the entangling capability, we implement the \enquote{Bell-Measurement}~\cite{foulds_controlled_2021,haug_scalable_2023,foulds_generalising_2024}.
We consider a circuit where the state of interest is vertically prepared twice and an inverse Bell-state is applied onto each pair of qubits between the two subsystems.
The setup is depicted in \autoref{fig:bell_measurement}.

It was shown that the squared trace of the density matrix is linearly dependent on the probability of measuring the parity between each of the qubits in the individual subsystems:

\begin{equation}
    \operatorname{Tr}\left[\rho_{k}^{2}\right]=1-2 \cdot P_{\text {odd}, k},
\end{equation}

where $P_{\text {odd}, k}$ is the probability of odd, non-zero parity in the outcomes of the $k$th qubit on each copy.
Inserting this in~\autoref{eq:entangling_capability_brennen}, translates to
\begin{equation}
    \left.Q(|\psi\rangle\right)=2\left(1-\frac{1}{n} \sum_{k=0}^{n-1}\left(1-2 \cdot P_{\mathrm{odd}, k}\right)\right),
\end{equation}
resulting in the same estimate of the entangling capability as when using the squared trace of the density matrix directly.

The \pythoninline{bell_measurements} method within the \pythoninline{Entanglement} in \qmlessentials provides a way to calculate the Bell-measurement, given a number of parameter samples.

While the computational costs of the \enquote{Bell-Measurement} also scale exponentially with the number of qubits in a classical simulation of the \ac{qml} model, it provides a physical observable measurement of the entangling capability, applicable to real quantum systems.

\subsection{Noise}

As long as \acp{ftqc} are not available, noise remains one of the primary challenges of \ac{qml}, not only distorting model predictions, but also limiting the trainability~\cite{ragone_lie_2024}.
As the effect of noise on \ac{qfm} is an interesting research direction~\cite{fontana_spectral_2022}, we offer a low-barrier interface to enable various types of noise as shown in~\autoref{tab:noises} to be added to a \ac{qml} model.

\begin{table}[htb]
    \centering
    \caption{Available noise types in \qmlessentials.}
    \label{tab:noises}
    \begin{tblr}{width=\linewidth,
            colspec={lX[l]},
            row{2-Z}={belowsep=0.15em},
            booktabs}
        \toprule
        \textbf{Noise}                        & \textbf{Description}                                                                                                                                  \\
        \midrule
        \pythoninlinetable{BitFlip}           &                                                                                                                                                       %Bit-Flip with specified probability $p_\text{bf}$ applied on each gate\\
        Per-gate bit-flip with probability $p_\text{bf}$.                                                                                                                                             \\
        %\hline
        \pythoninlinetable{PhaseFlip}         & Per-gate phase-flip with probability $p_\text{pf}$.                                                                                                   \\
        %\hline
        \pythoninlinetable{Depolarization}    & Per-gate depolarisation with probability $p_\text{dp}$.                                                                                               \\
        %\hline
        \pythoninlinetable{AmplitudeDamping}  & Amplitude Damping noise with probability $p_\text{ad}$ applied at the end of the circuit.                                                             \\
        %\hline
        \pythoninlinetable{PhaseDamping}      & Phase Damping noise with probability $p_\text{pd}$ applied at the end of the circuit.                                                                 \\
        %\hline
        \pythoninlinetable{ThermalRelaxation} & Thermal relaxation of the system characterised by \pythoninlinetable{t1}, \pythoninlinetable{t2}  and gate time factor \pythoninlinetable{t\_factor}. \\
        %\hline
        \pythoninlinetable{Measurement}       & Bit-Flip error with probability $p_\text{me}$ applied at the very end of the circuit.                                                                 \\
        %\hline
        \pythoninlinetable{StatePreparation}  & Bit-Flip error with probability $p_\text{sp}$ applied at the very beginning of the circuit.                                                           \\
        %\hline
        \pythoninlinetable{GateError}         & Imprecise gate operations with error $\epsilon \sim \mathcal{N}(0, \mu)$ as coherent per-gate noise.                                                  \\
        \bottomrule
    \end{tblr}
\end{table}

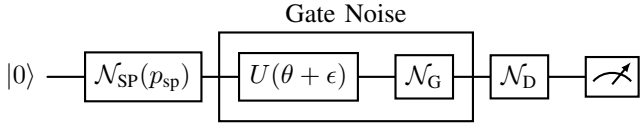
\begin{figure}[htb]
    \centering
    \begin{quantikz}
        \lstick{\ket{0}} & \gate{\mathcal{N}_\text{SP}(p_\text{sp})} & \gate{U(\theta + \epsilon)}\gategroup[wires=1,steps=2]{Gate Noise} & \gate{\mathcal{N}_\text{G}} & \gate{\mathcal{N}_\text{D}} &  \meter{}
    \end{quantikz}
    \caption{Noise model consisting of state preparation noise $\mathcal{N}_\text{SP}$, coherent rotational error $\epsilon$ that goes into a noisy gate $U$ with incoherent gate error $\mathcal{N}_\text{G}$, damping noise $\mathcal{N}_D$.}
    \label{fig:noise_model_all}
\end{figure}

Our parameterisable noise model is depicted in \autoref{fig:noise_model_all} for a single qubit model.

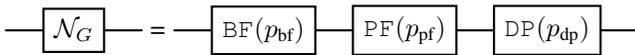
\begin{figure}[htb]
    \centering
    \begin{quantikz}
        & \gate{\mathcal{N}_G} &
    \end{quantikz}=\begin{quantikz}
        & \gate{\texttt{BF}(p_\text{bf})} & \gate{\texttt{PF}(p_\text{pf})} & \gate{\texttt{DP}(p_\text{dp})} &
    \end{quantikz}
    \caption{Decomposition of a single noise operation applied after each gate into bit-flip, phase-flip and depolarising noise.}
    \label{fig:noise_model_gate}
\end{figure}

\begin{figure}[htb]
    \centering
    \begin{quantikz}
        & \gate{\mathcal{N}_D} &
    \end{quantikz}=\begin{quantikz}
        & \gate{\texttt{AD}(p_\text{ad})} & \gate{\texttt{PD}(p_\text{pd})} & \gate{\texttt{BF}(p_\text{me})} &
    \end{quantikz}
    \caption{Decomposition of a noise operation applied at the end of the circuit into amplitude- and phase damping noise.}
    \label{fig:noise_model_end}
\end{figure}
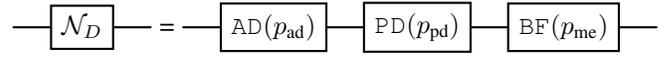

\autoref{fig:noise_model_gate} shows the decomposition of the gate noise $\mathcal{N}_\text{G}$ into bit-flip, phase-flip and depolarisation.
Similarly,~\autoref{fig:noise_model_end} shows the decomposition of the incoherent gate error $\mathcal{N}_\text{D}$ into amplitude damping and phase damping noise, as well as the measurement error, each of which is parametrised by its corresponding probability.

The individual types of incoherent noise, summarised in~\autoref{tab:noises} utilise the Kraus-operator mechanism implemented in PennyLane~\cite{bergholm_pennylane_2022}.
As \pythoninline{StatePreparation} and \pythoninline{Measurement} errors are not directly implemented in PennyLane, we model these by applying a bit-flip Kraus channel at the beginning, or end of the circuit, respectively.
Additionally, the coherent \pythoninline{GateError}, in which $\epsilon \sim \mathcal{N}(0, \mu)$ is drawn randomly from a Gaussian distribution for each gate, allows for modelling unintended rotation offsets on potential real devices.

For specific noise parameter tuning to represent actual quantum systems, we refer to Ref.~\cite{georgopoulos_modeling_2021} for a review.
By providing a \pythoninline{noise_params} Python dictionary to the call of the \pythoninline{Model} class, the full parametrised noise model can be applied to the corresponding \ac{vqc}.

\subsection{Ansätze}

\qmlessentials includes a set of predefined Ansätze that can be used to build up a model.
Currently, this set includes
\begin{itemize*}
    \item Circuit 1*,
    \item Circuit 2*,
    \item Circuit 3*,
    \item Circuit 4*,
    \item Circuit 6*,
    \item Circuit 9*,
    \item Circuit 10*,
    \item Circuit 15*,
    \item Circuit 16*,
    \item Circuit 17*,
    \item Circuit 18*,
    \item Circuit 19*,
    \item No Entangling,
    \item Strongly Entangling,
    \item Hardware Efficient,
\end{itemize*}.
Ansätze marked with * are implemented based on the Work from Sim~\etal~\cite{sim_expressibility_2019} and can be viewed in their corresponding paper.

\begin{figure}
    \centering
    \begin{tikzpicture}
        \node[scale=0.85] {
            \begin{quantikz}[row sep=2mm, column sep=2mm]
                & \gate{R_Y(\theta^0_0)} & \gate{R_Z(\theta^1_0)} & \gate{R_Y(\theta^2_0)} & \ctrl{1} & & \targ{} & \\
                & \gate{R_Y(\theta^0_1)} & \gate{R_Z(\theta^1_1)} & \gate{R_Y(\theta^2_1)} & \targ{} & \ctrl{1} & & \\
                & \gate{R_Y(\theta^0_3)} & \gate{R_Z(\theta^1_3)} & \gate{R_Y(\theta^2_3)} & \ctrl{1} & \targ{} & & \\
                & \gate{R_Y(\theta^0_4)} & \gate{R_Z(\theta^1_4)} & \gate{R_Y(\theta^2_4)} & \targ{} & & \ctrl{-3} &
            \end{quantikz}
        };
    \end{tikzpicture}
    \caption{A single layer of a circular $4$-qubit Hardware Efficient Ansatz.}
    \label{fig:hw_eff_ansatz}
\end{figure}
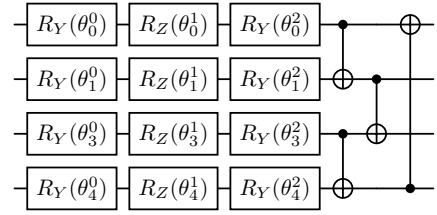

For the Hardware-Efficient Ansatz, which only utilises native gates of a typical quantum device and linearly scaling circuit depths, we chose the structure as shown in \autoref{fig:hw_eff_ansatz}.
We acknowledge that this approach is sometimes realised without the last \texttt{CNOT} gate, which we included for symmetric reasons.

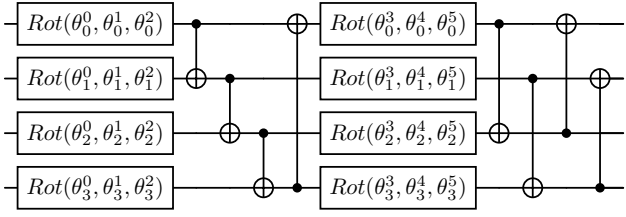
\begin{figure}
    \centering
    \begin{tikzpicture}
        \node[scale=0.85] {
            \begin{quantikz}[row sep=2mm, column sep=2mm]
                \qw & \gate{Rot(\theta^0_0, \theta^1_0, \theta^2_0)} & \ctrl{1} &  &  & \targ{} & \gate{Rot(\theta^3_0, \theta^4_0, \theta^5_0)} & \ctrl{2} &  & \targ{} & &    \qw \\
                \qw & \gate{Rot(\theta^0_1, \theta^1_1, \theta^2_1)} & \targ{} & \ctrl{1} &  &  & \gate{Rot(\theta^3_1, \theta^4_1, \theta^5_1)} &  & \ctrl{2} &  & \targ{} &   \qw \\
                \qw & \gate{Rot(\theta^0_2, \theta^1_2, \theta^2_2)} &  & \targ{} & \ctrl{1} &  & \gate{Rot(\theta^3_2, \theta^4_2, \theta^5_2)} & \targ{} &  & \ctrl{-2} & &   \qw \\
                \qw & \gate{Rot(\theta^0_3, \theta^1_3, \theta^2_3)} &  &  & \targ{} & \ctrl{-3} & \gate{Rot(\theta^3_3, \theta^4_3, \theta^5_3)} &  & \targ{} &  & \ctrl{-2} & \qw
            \end{quantikz}
        };
    \end{tikzpicture}
    \caption{A single layer of a $4$-qubit Strongly Entangling Ansatz as introduced in Ref.~\cite{schuld_circuit-centric_2020}.}
    \label{fig:strongly_entangling_ansatz}
\end{figure}

The Strongly-Entangling Ansatz is inspired by~\cite{schuld_circuit-centric_2020} and displayed in \autoref{fig:strongly_entangling_ansatz}.

As data embedding we utilise a Pauli $R_X$ rotation as default.
Further embeddings and Ansätze can be added by providing a callable function upon instantiation of the model.

\section{Examples and Validation}
\label{sec:examples}

In this section, numerical results are presented that validate the implementation of the coefficients, expressibility and entanglement calculation using a subset of the Ansätze presented in Ref.~\cite{sim_expressibility_2019}.

Notable, the results presented here show only a fraction of what is covered by automated testing in our continuous integration and development pipeline.
For all of the following numerical results we use a model with $4$ qubits and a single variational layer if not stated otherwise.
The parameters of the model are sampled from a uniform distribution between $0$ and $2\pi$, with $200$ samples in the case of the coefficients and $5000$ samples for the expressibility and entanglement calculations.

\subsection{Coefficients}

\begin{figure}[htb]
    \centering
    \includegraphics[width=0.48\textwidth]{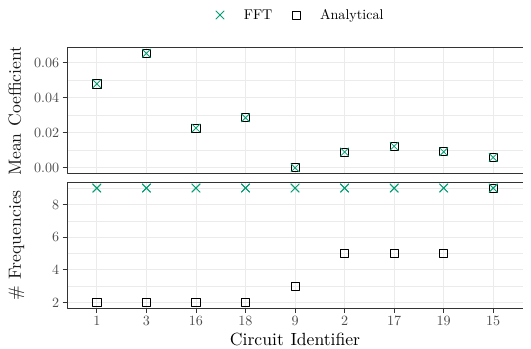}
    \caption{Top: Fourier coefficients mean values of a circuit obtained using the analytical method and the \ac{fft}. Bottom: Difference between the two methods in terms of the number of frequencies.}
    \label{fig:coefficients_valid}
\end{figure}

Although no reference results are available for the circuits introduced in Ref.~\cite{sim_expressibility_2019}, we can validate our implementation of the coefficients by comparing the analytical calculation with the results obtained using the \ac{fft}.
In this experiment we focus on the measurement of a single qubit.
The upper part of \autoref{fig:coefficients_valid} shows the results of this validation, where each measurement represents the average of the coefficients over all frequencies.
As mentioned in \autoref{par:coefficients_fft}, the calculation of coefficients using the \ac{fft} bears some caveats when it comes to the estimation of the correct number of frequencies.
We demonstrate this discrepancy in the lower part of \autoref{fig:coefficients_valid} where it is clearly visible that only a few circuits actually contain the frequencies indicated by the \ac{fft}.
By filtering out the frequencies that are not present in the circuit, we obtain identical results from both methods.

\subsection{Expressibility}

\begin{figure}[htb]
    \centering
    \includegraphics[width=0.48\textwidth]{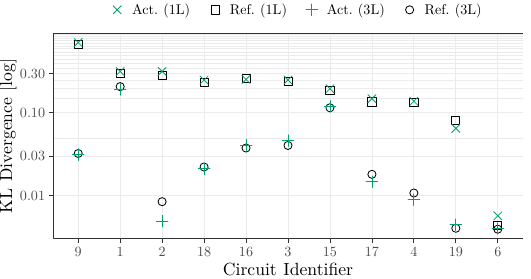}
    \caption{Expressibility of various circuit with respect to the Haar distribution as \ac{kl} divergence as reference from~\cite{sim_expressibility_2019} (black) and our implementation (teal) for one and three layers.}
    \label{fig:expressibility_valid}
\end{figure}

The validation of our implementation of the expressibility is performed by referring to the results from the study by Sim~\etal~\cite{sim_expressibility_2019}.
\autoref{fig:expressibility_valid} presents the outcomes of this validation, with expressibility defined as the inverse of the \ac{kl} divergence between the Haar distribution and the model distribution.
It is evident that there is a near-perfect alignment between the reference results and the experimental findings.

\subsection{Entangling Capability}

\begin{figure}[htb]
    \centering
    \includegraphics[width=0.48\textwidth]{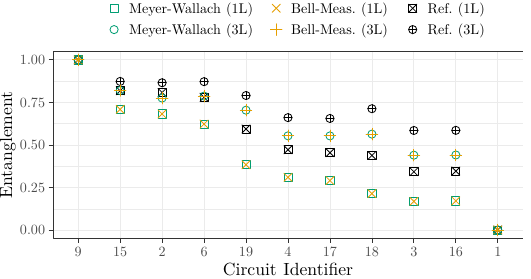}
    \caption{Entangling Capability of various circuits as reference from~\cite{sim_expressibility_2019} (black) and our implementation using the Meyer-Wallach (teal) measure and Bell-Measurements (orange).}
    \label{fig:entanglement_valid}
\end{figure}

Similarly to expressibility, we validate our implementation of calculating the entangling capability using the results from Ref.~\cite{sim_expressibility_2019} as a reference with the exact same setup.
\autoref{fig:entanglement_valid} shows the results of this validation.
It can be observed that the entangling capability is generally overestimated with an increasing discrepancy towards lower values as well as a higher number of layers.
The discrepancy may be partly attributed to the unavailability of the precise results from the experiments in Ref.~\cite{sim_expressibility_2019}.
Furthermore, the present implementation is based on the work of Brennen~\etal~\cite{brennen_observable_2003}.
While this is technically equivalent, it is not exactly the same as the method used in Sim~\etal~\cite{sim_expressibility_2019}.
However, it is notable that the results of both methods, the Meyer-Wallach and the Bell-Measurements, align perfectly.

\section{Conclusion and Outlook}
\label{sec:conclusion}

With \qmlessentials we provide a framework that aims to give researchers a tool at hand to explore properties of \acp{qfm}.
To this end, our framework provides algorithms to calculate the expressibility and entangling capability as well as the Fourier coefficients of a provided \acp{qfm}.
The latter can be achieved by either a \ac{fft} or analytical method with the advantage of obtaining the \enquote{true} number of coefficients.
We implement various tests with the intention of ensuring the accuracy of provided algorithms and aim for a maximum of flexibility to allow adaptation to a variety of use cases.

In future work, we want to extend our framework by adding more testing and validation while also implementing more features including
\begin{enumerate*}[label=(\arabic*)]
    \item pulse-level control of \acp{qfm},
    \item additional measures for entanglement and expressibility,
    \item trainable frequencies in the encoding unitaries~\cite{jaderberg_let_2024},
    \item support for the Qiskit~\cite{javadi-abhari_quantum_2024} framework and
    \item multiprocessing capabilities
\end{enumerate*}.

\section*{Acknowledgment}
\blackout{MS}, \blackout{EK} and \blackout{AS} acknowledge support by \blackout{the state of Baden-W\"urttemberg through bwHPC}.
\blackout{MF} and \blackout{WM} acknowledge support by \blackout{the German Federal Ministry of Education and Research (BMBF)}, funding program \blackout{\enquote{Quantum Technologies—--From Basic Research to Market}}, grant number \blackout{13N16092}.
\blackout{WM} acknowledges support by the \blackout{High-Tech Agenda of the Free State of Bavaria}.
We would like to express our gratitude for the contributions in the form of issues, pull requests and code to \qmlessentials by \blackout{Paul Schillinger}, \blackout{Johannes Wahl} and \blackout{Clotilde Guyard-Gilles}.

\printbibliography

\end{document}